\documentclass[conference]{IEEEtran}
\IEEEoverridecommandlockouts
\pdfoutput=1 
\usepackage{cite}
\usepackage{amsmath,amssymb,amsfonts}
\usepackage{algorithmic}
\usepackage{graphicx}
\usepackage{textcomp}
\usepackage{xcolor}
\def\BibTeX{{\rm B\kern-.05em{\sc i\kern-.025em b}\kern-.08em
    T\kern-.1667em\lower.7ex\hbox{E}\kern-.125emX}}
\usepackage{color, colortbl}
\usepackage[ruled,vlined]{algorithm2e}

\ifCLASSOPTIONcompsoc
\usepackage[caption=false,font=normalsize,labelfon
t=sf,textfont=sf]{subfig}
\else
\usepackage[caption=false,font=footnotesize]{subfi
g}
\fi
\def\BibTeX{{\rm B\kern-.05em{\sc i\kern-.025em b}\kern-.08em T\kern-.1667em\lower.7ex\hbox{E}\kern-.125emX}}
\definecolor{Gray}{gray}{0.9}   
\graphicspath{{./figs/}}
\begin{document}

\title{Reinforcement Learning-based Admission Control \\in Delay-sensitive Service Systems \\
}
\author{\IEEEauthorblockN{Majid Raeis, Ali Tizghadam and Alberto Leon-Garcia}
\IEEEauthorblockA{\textit{Department of Electrical and Computer Engineering} \\
\textit{University of Toronto}, Toronto, Canada \\
Emails: m.raeis@mail.utoronto.ca, ali.tizghadam@utoronto.ca  and alberto.leongarcia@utoronto.ca}
}

\maketitle

\begin{abstract}
Ensuring quality of service (QoS) guarantees in service systems is a challenging task, particularly when the system is composed of more fine-grained services, such as service function chains.
An important QoS metric in service systems is the end-to-end delay, which becomes even more important in delay-sensitive applications, where the jobs must be completed within a time deadline. Admission control is one way of providing end-to-end delay guarantee, where the controller accepts a job only if it has a high probability of meeting the deadline. In this paper, we propose a reinforcement learning-based admission controller that guarantees a probabilistic upper-bound on the end-to-end delay of the service system, while minimizes the probability of unnecessary rejections. Our controller only uses the queue length information of the network and requires no knowledge about the network topology or system parameters. Since long-term performance metrics are of great importance in service systems, we take an average-reward reinforcement learning approach, which is well suited to infinite horizon problems. Our evaluations verify that the proposed RL-based admission controller is capable of providing probabilistic bounds on the end-to-end delay of the network, without using system model information.
\end{abstract}

\begin{IEEEkeywords}
Admission control, queueing networks, reinforcement learning, delay-sensitive applications.
\end{IEEEkeywords}
\vspace{-0.2cm}
\section{introduction}\label{intro}

Providing quality of service (QoS) guarantees in complex service systems is often a challenging task. Service function chaining (SFC) is one such example, in which the end-to-end service is provided through a sequence of service functions (SFs) such as firewalls, load balancers and deep packet inspectors. One important QoS metric in service chains is the end-to-end delay, which is particularly important in delay-sensitive applications in which a job must be completed within some specific deadline. Admission control (AC) is one way of providing QoS, where the controller offers end-to-end delay guarantees by rejecting those that are likely to fail the delay requirement.  This may also result in a higher throughput, since there will be more room for the future arrivals.

An important challenge in designing controllers for service networks is the lack of knowledge about the dynamics of the system, particularly when the system becomes more complex. This is a reason why classic network control algorithms often fall short on practicality. Reinforcement learning is a natural candidate that can deal with this issue. In the RL framework, the agent (controller) interacts with the environment (network system) and optimizes its policy without knowledge about the dynamics or topology of the system. In this paper we focus on RL-based control mechanisms for providing QoS in service systems, without limiting ourselves to a particular application.

One of the earliest works on the use of reinforcement learning for call admission control is \cite{brown}, which aims to maximize the earned revenue while providing QoS guarantees to the users. An RL-based call admission controller for cellular networks has been proposed in~\cite{senouci}, which  improves the quality of service and reduces call-blocking probabilities of hand-off calls.
Service function chaining is another example of service networks, in which QoS can be of great importance.
The authors in~\cite{5g} proposed a DQN (Deep Q-Learning) based QoS/QoE aware service function chaining in NFV-enabled 5G networks. This work considered QoS metrics such as delay, throughput, bandwidth, etc. The authors of~\cite{autosac1,autosac2} proposed an automatic service and admission controller, called AutoSAC, with an application in network function chaining. AutoSAC provides a service controller for automatic VNF scaling, and an admission controller that guarantees that the accepted jobs meet their end-to-end deadlines. The proposed admission controller uses the worst-case expected delays to make its decision, which involve loose estimates of the delay and therefore jeopardise the throughput. In~\cite{iot}, the authors proposed a deep reinforcement learning approach to handle complex and dynamic SFC embedding scenarios in IoT, where the average SFC processing delay has been used as the primary embedding objective. Another body of literature studies the problem of QoS measurement and control in the context of general queueing systems. The end-to-end delay distribution of the tandem and acyclic queueing networks have been studied in~\cite{majid1,majid2}, where mixture density networks (MDNs)~\cite{sutton} have been used to learn the distributions. The distributions are used for providing probabilistic bounds on the end-to-end delay of the network. The authors in~\cite{liu} have used a general queueing model to learn a network control policy. More specifically, a model-based reinforcement learning approach has been used to find the optimal policy that minimizes the average job delay in queueing networks. The reader can refer to~\cite{datacenter,survey,mul_agent} for some related works in other service system contexts.

In this paper, we use queueing theory as a general framework for modeling service systems, and therefore our results are applicable to a wide range of applications. In contrast to the prior works, we adopt an average-reward reinforcement learning approach, which is well-suited for non-episodic tasks where long-term performance metrics are of great importance~\cite{sutton, park}. Our main contributions can be summarized as follows
\begin{itemize}
    \item We propose an RL-based admission controller that provides a probabilistic upper-bound on the end-to-end delay of the system, while most of the existing work focus on the average delay, which is less informative.
    
    \item Our controller is able to minimize the probability of unnecessary rejections. In order to accomplish this, we use a simulated environment in parallel with the original environment to determine whether a rejection decision was a good decision, or a wrong choice and that the job could have been accepted.
    
    \item In contrast to the classical queueing methods, our controller only observes the queue length information and does not require any information about the network topology, service or inter-arrival time distributions. 
    
\end{itemize}

The paper is organized as follows. In Section~\ref{sys_model}, we describe the queueing system model and formulate the admission control problem as an optimization problem. We briefly review some basic concepts in reinforcement learning, especially the average-reward setting, in Section~\ref{RLintro}. In Section~\ref{AC_RL}, we formulate the problem as an average-reward reinforcement learning and discuss the implementation challenges that need to be addressed. The evaluation of the proposed RL-based admission controller is presented in the context of service function chaining in Section~\ref{eval}. Finally, Section~\ref{con} presents conclusions.

\section{System Model and Problem Setting}\label{sys_model}
\begin{figure}[!t] 
\centering
\subfloat[]{\includegraphics[trim={5cm 9cm 5cm 9cm},clip, scale=0.42]{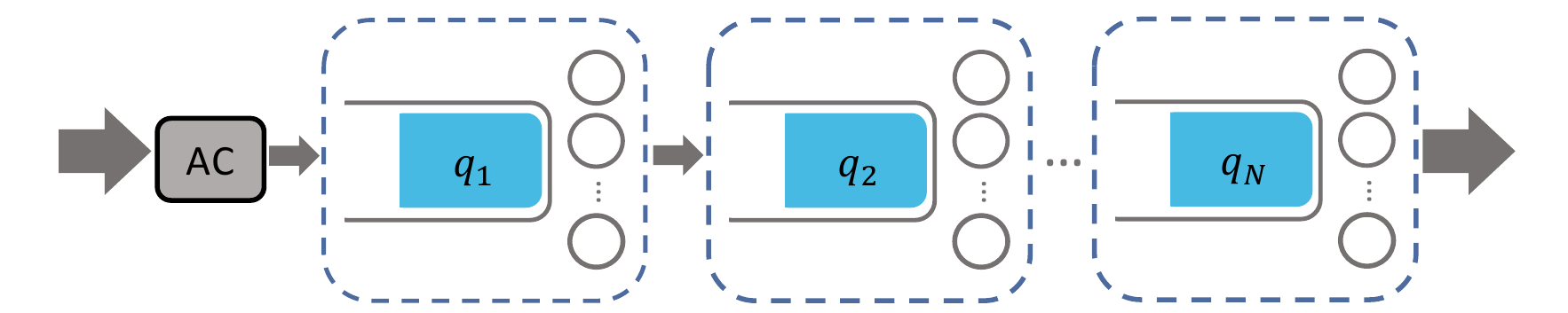}
\label{fig:tandem_topo}}
\hfill
\subfloat[]{\includegraphics[trim={5cm 6.5cm 5cm 6.5cm},clip,scale=0.42]{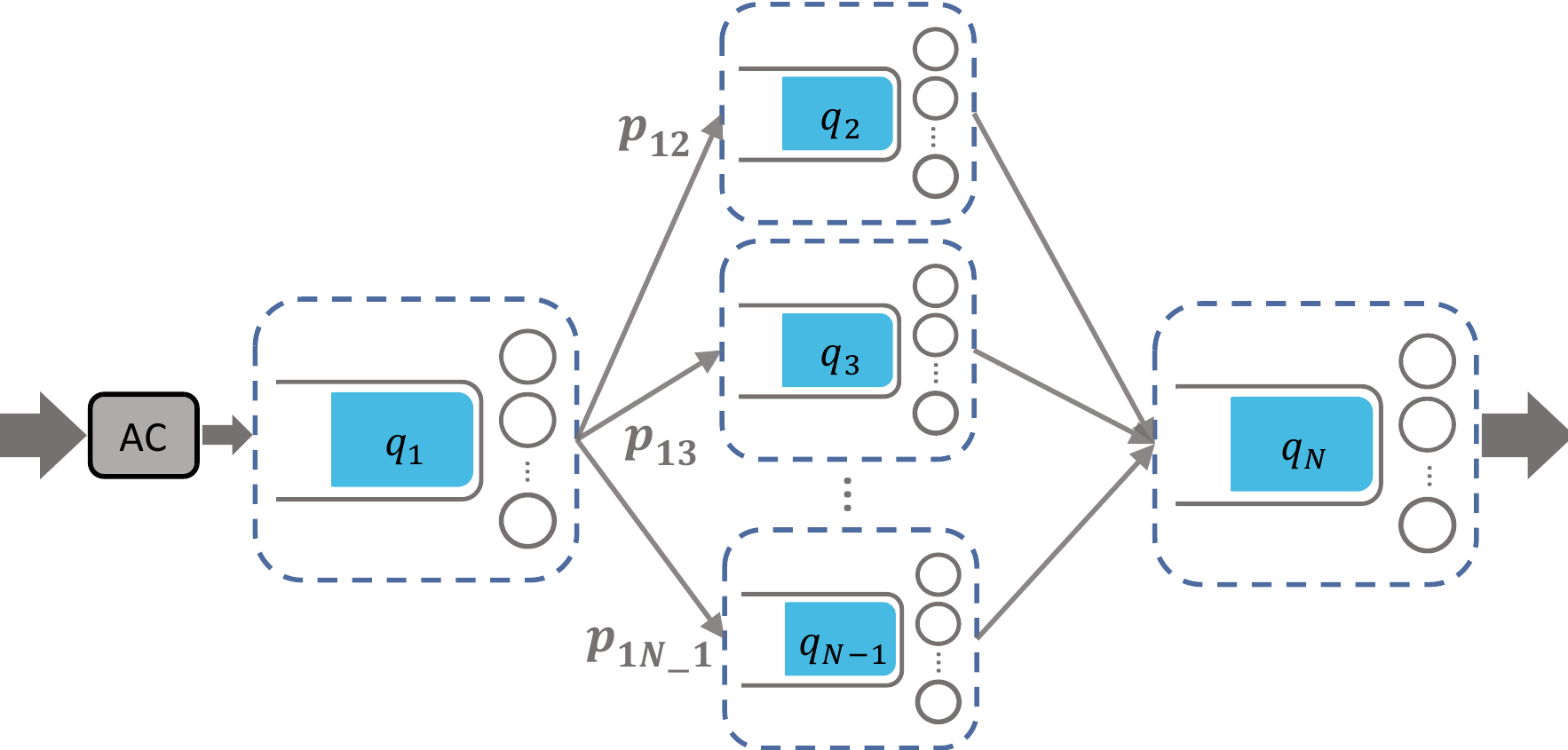}
\label{fig:acyclic_topo}}
\caption{Network topologies (a) Tandem queue (b) Acyclic queue.}
\label{fig:topology}
\vspace{-0.5cm}
\end{figure}

We consider multi-server queueing systems, with First Come First Serve (FCFS) service discipline, as the building blocks of the service networks that we study. Furthermore, we study tandem queues and simple acyclic queueing networks as shown in Fig.\ref{fig:topology}. In a tandem topology, a customer must go through all the stages to receive the end-to-end service, while in an acyclic topology, the customers randomly go through one of the branches with the specified probabilities in Fig.~\ref{fig:acyclic_topo}. We do not assume specific distributions for the service times
or the inter-arrival times and therefore, these processes can have arbitrary stationary distributions.

Now, consider a network consisting of $N$ queueing systems, where system $n$, $1\leq n \leq N$, is a multi-server queueing system with $c_n$ homogeneous servers having service rates $\mu_n$. Let $q_n$ denote the queue length of the $n$th queueing system upon arrival of a job at the entrance of the network ($1$st queue). The end-to-end delay of a new arrival is represented by $d$. Moreover, we consider an admission controller at the entrance of the network, which decides whether to accept or reject an incoming job based on the queue length information of all the constituent systems, i.e., $(q_1,q_2,\cdots,q_N)$. The policy of the admission controller is represented by $\pi$, and the acceptance and rejection actions are denoted by $A$ and $R$, respectively.

As mentioned earlier, the reason for considering the admission controller is to provide some sort of QoS to the customers. More specifically, our goal in designing the admission controller is to guarantee a probabilistic upper-bound on the end-to-end delay of the accepted jobs, i.e., $P(d>d_{ub}|A)\leq \epsilon_{ub}$, where $d_{ub}$ denotes the upper-bound and $\epsilon_{ub}$ is the violation probability. Many different policies may result in the same probabilistic upper-bound, and we are interested in the policy that results in the minimum probability of unnecessary rejections. Therefore, we can express the admission control design as an optimization problem:
\begin{align} \label{Eq:opt1}
    \max_{\pi}&\qquad-p_\pi^a(R)P(d<d_{ub}|R) \nonumber \\
    \text{s.t.}&\qquad P(d<d_{ub}|A)\geq 1-\epsilon_{ub},
\end{align}
where $p_\pi^a(a)$, $a\in\{A,R\}$, is the steady state probability of choosing action $a$ under policy $\pi$, and the objective is to minimize the probability of unnecessary rejection of an arrival that could have met the deadline, i.e., $P\big(R\cap (d<d_{ub})\big)=p_\pi^a(R)P(d<d_{ub}|R)$.

\section{Background on Reinforcement Learning}\label{RLintro}

In this section, we briefly review the basic concepts of the reinforcement learning and discuss the average-reward setting.
The basic elements of a reinforcement learning problem are the \emph{agent} and the \emph{environment}, which have iterative interactions with each other. The environment is modeled by a Markov decision process (MDP), which is specified by $<\mathcal{S},\mathcal{A},\mathcal{P},\mathcal{R}>$, with state space $\mathcal{S}$, action space $\mathcal{A}$, state transition probability matrix $\mathcal{P}$ and reward function $\mathcal{R}$. At each time step $t$, the agent observes a state $s_t \in \mathcal{S}$, takes action $a_t \in \mathcal{A}$, transits to state $s_{t+1} \in \mathcal{S}$ and receives a reward of $\mathcal{R}(s_t,a_t,s_{t+1})$. The agent's actions are defined by its policy $\pi$, where $\pi(a|s)$ is the  probability of taking action $a$ in state~$s$.

The formal definition of an agent's goal varies based on the problem setting. In episodic tasks in which there is a notion of terminal state, the goal of the agent is to maximize the long-term expected \emph{return}, where the return is defined as $G_t = \sum_{k=t+1}^T R_k
$. In continuing settings, in which the interaction between the agent and the environment continues forever, we cannot use the same return function as before, since $T=\infty$ and the reward can easily become unbounded. One way of handling this issue is by using \emph{discounting}. In this approach, the agent tries to maximize the expected discounted return, which is defined as $G_t = \sum_{k=t+1}^T \gamma^{k-t-1} R_k$, $0\leq \gamma<1$. In the discounted setting, the agent cares more about the immediate rewards in the near future rather than the delayed rewards in the far future. However, in many applications such as computer networks, we are interested in the average performance of the system in the long run and we care as much about the future as we do about the present. There is a third setting for formulating the goal in RL problems, which is called the \emph{average reward} setting\cite{sutton}. In this setting, the goal of the agent is to maximize the average reward per time step, which under \emph{ergodicity} assumption can be obtained as~\cite{sutton}
\begin{equation}
    r(\pi) = \sum_s p_{\pi}^s(s) \sum_a \pi(a|s) \sum_{s',r} P(s',r|s,a)r,
\end{equation}
where $p_{\pi}^s(s)$ is the steady state distribution of being in state $s$ in a given time step following policy $\pi$. The average reward setting is a good candidate for formulating the goal in continuing environments, where we care about the future as much as the present. The \emph{differential} return is defined as $G_t = \sum_{k=t+1}^\infty (R_k-r(\pi))$. Similar to the notion of action-value function in the discounted setting, we can define differential action-value function (Q-function) as  $Q_\pi(s,a)=\mathbb{E}_\pi[G_t|s_t=s,a_t=a]$, which denotes the expected return starting from state $s$, taking action $a$, and following policy $\pi$. Defining the optimal Q-function as $Q^*(s,a)=\max_{\pi}Q_{\pi}(s,a)$, we can obtain the optimal policy as
\begin{equation}
\pi^*(a|s) = \left\{
\begin{array}{ll}
1\qquad \text{if} & a={\arg \max}_{a'} Q^*(s,a'),\\
0\qquad & \text{otherwise}.
\end{array} \right.
\end{equation}
The optimal Q-function must satisfy the Bellman optimality equation for the average reward setting as follows 
\begin{equation}
    Q^*(s,a)=\mathbb{E}_\pi\bigg[r-\max_\pi r(\pi)+\max_{a'} Q^*(s',a')\Big|s,a\bigg].
\end{equation}
Similar to the discounted setting, the Bellman optimality equation is not usually directly used to obtain the optimal Q-function. Instead, we use an iterative method to solve the Bellman equation, which is similar to the well-known Q-learning algorithm and is called \emph{R-learning}~\cite{rlearning}. Based on this method, the Q-function is updated at each time step as
\begin{equation}
    Q(s_t,a_t)\hspace{-0.1cm}\leftarrow \hspace{-0.1cm} Q(s_t,a_t)+\alpha \hspace{-0.06cm} \big[r_{t+1}-\bar{r}+\max_a Q(s_{t+1},a)-Q(s_t,a_t\hspace{-0.06cm})\big]\hspace{-0.08cm},
\end{equation}
where $\alpha$ represents the learning rate and $\bar{r}$ is an approximation of $r(\pi)$. At each time step that the behaviour policy acts greedily, i.e., $a_t=\arg\max_a Q(s_t,a)$, $\bar{r}$ is updated as
\begin{equation}
    \bar{r}\leftarrow \bar{r}+\beta \big[r_{t+1}-\bar{r}+\max_a Q(s_{t+1},a)-Q(s_t,a_t)\big],
\end{equation}
where $\beta$ is the step-size parameter.
\section{Admission Control as a Reinforcement Learning Problem}\label{AC_RL}

\subsection{Problem Formulation}
In this section, we formulate the admission control task as a reinforcement learning problem in the average-reward setting. Our environment is a tandem (acyclic) queueing network and the agent is the admission controller at the entrance of the network. The goal is to design an admission controller that minimizes the average number of unnecessary rejections per time step, while guaranteeing a probabilistic upper-bound on the end-to-end delay of the network.
In order to achieve this goal, our controller can interact with the environment upon arrival of each job. Therefore, each time step is the interval between two consecutive job arrivals.

Now, let us define the components of our reinforcement learning problem as follows:
\begin{itemize}
    \item \textbf{State:} The vector of queue lengths of all the constituent queueing systems upon a job arrival, i.e., $s=(q_1, q_2, \cdots, q_N)$, where $q_n$ represents the queue length of the $n$th queueing system. 
    \item \textbf{Action:} The possible actions are whether to accept ($a=A$) or reject ($a=R$) a job upon its arrival. Here we consider deterministic policies and therefore, action will be a deterministic function of the state, i.e. $\pi(a|s)=0~\text{or}~1, a~\in~\{A, R\}$. Therefore, we use $a(s)=i, i\in\{A, R\}$ to show the taken action at state $s$.
    \item \textbf{Reward:} Designing the reward function is the most challenging part of the problem. The reward function must be defined such that maximizing the average reward results in the desired goal formulated in (\ref{Eq:opt1}).
    In order to simplify the discussion, we make the following assumption: we assume that we know the end-to-end delay of the network at any time $t$, i.e., the end-to-end delay that a potential arrival at time $t$ would experience if it is accepted by the admission controller. This is clearly an unrealistic assumption, since some of the accepted jobs might not be departed by the next time step and some might be even rejected and therefore never go through the network. However, this assumption will help us in designing the reward function. We will discuss how the reward function can be calculated under realistic assumptions later in this section. 
    
    Now, consider the following reward function:
    \begin{equation}\label{Eq:reward}
    r_n = \left\{
    \begin{array}{ll}
    r^A_1\qquad \text{if} & a = A \text{ and } d_n<d_{ub},\\
    r^A_2\qquad \text{if} & a = A \text{ and } d_n>d_{ub},\\
    r^R_1\qquad \text{if} & a = R \text{ and } d_n<d_{ub},\\
    r^R_2\qquad \text{if} & a = R \text{ and } d_n>d_{ub},
    \end{array} \right.
    \end{equation}
    where $r_n$ denotes the immediate reward obtained in time step $n$, and $d_n$ represents the end-to-end delay of the $n$th arrival.
    The average reward per time step for policy $\pi$ can be written as
    \begin{align}  \label{Eq:ave_rew1}
        r(\pi) &= \sum_s p_\pi^s(s)\mathbb{E}_\pi[r|s] \nonumber \\ &=\hspace{-0.25cm}\sum_{\substack{s~\text{s.t.}\\a(s)=A}}p_\pi^s(s)\mathbb{E}_\pi[r|s] +\hspace{-0.25cm} \sum_{\substack{s~\text{s.t.}\\a(s)=R}}p_\pi^s(s)\mathbb{E}_\pi[r|s] \nonumber \\
        &=\hspace{-0.25cm} \sum_{\substack{s~\text{s.t.}\\a(s)=A}}\hspace{-0.2cm}p_\pi^s(s)\left[r_1^A P(d<d_{ub}|s)+r_2^A P(d>d_{ub}|s)\right] \nonumber \\
        &+ \hspace{-0.25cm}\sum_{\substack{s~\text{s.t.}\\a(s)=R}}\hspace{-0.2cm}p_\pi^s(s)\left[r_1^R P(d<d_{ub}|s)+r_2^R P(d>d_{ub}|s)\right]\hspace{-0.1cm}, \nonumber \\
    \end{align}
    where $P(d<d_{ub}|s,\pi)$ is shown by $P(d<d_{ub}|s)$ for simplicity. Furthermore, since for a given action $a\in\{A, R\}$, we have
    \begin{align*}
    \sum_{\substack{s~\text{s.t.}\\a(s)=a}} p_\pi^s(s)P(d<d_{ub}|s)=&P((d<d_{ub})\cap a)\\ =& p_\pi^a(a) P(d<d_{ub}|a),
    \end{align*}
    where $p_{\pi}^a(a)=\sum_{s}p_\pi^s(s)\pi(a|s)$, we can write Eq.~(\ref{Eq:ave_rew1}) as
    \begin{align}\label{eq:ave_rew2}
        r(\pi) &= p_\pi^a(A) \left[(r_1^A-r_2^A) P(d<d_{ub}|A)+r_2^A\right] \nonumber \\
        &+ p_\pi^a(R) \left[(r_1^R-r_2^R) P(d<d_{ub}|R) +r_2^R\right] .
    \end{align}
    Now, we should choose the parameters of the reward function such that the formulated goal in (\ref{Eq:opt1}) is achieved.

    Let us first define the Lagrangian function associated with problem~(\ref{Eq:opt1}) as
    \begin{align} \label{Eq:lag}
        L(\pi,\lambda) &= -p_\pi^a(R)P(d<d_{ub}|R) \nonumber \\
        &+ \lambda \left(P(d<d_{ub}|A)- (1-\epsilon)\right),
    \end{align}
    where $\lambda$ is the Lagrange multiplier associated with the QoS constraint in our optimization problem. The Lagrangian dual function is defined as $g(\lambda) = \max_{\pi} L(\pi, \lambda)$ and therefore the dual problem is
    \begin{equation}\label{eq:dual}
        \min_{\lambda} g(\lambda), \qquad \text{s.t. }\lambda\geq 0.
    \end{equation}
    Now, using Eq.~(\ref{eq:ave_rew2}) and choosing $r_1^A = \epsilon \lambda$, $r_2^A= -(1-\epsilon)\lambda$, $r_1^R = -1$ and $r_2^R = 0$, we have
    \begin{align}\label{eq:r_l_pi}
        r_{\lambda}(\pi) &= -p_\pi^a(R)P(d<d_{ub}|R) \nonumber \\
        &+ \lambda p_\pi^a(A)\left(P(d<d_{ub}|A)- (1-\epsilon)\right),
    \end{align}
    where we use $r_{\lambda}(\pi)$ instead of $r(\pi)$ to emphasize on its dependence on $\lambda$.
    As can be seen from Eqs.~(\ref{Eq:lag}) and ~(\ref{eq:r_l_pi}), $r_{\lambda}(\pi)$ corresponds to the Lagrangian function of the same problem as (\ref{Eq:opt1}), where both sides of the  inequality constraint are multiplied by $p_\pi^a(A)$, i.e.,
    \begin{align} \label{Eq:opt2}
        \max_{\pi}&\qquad-p_\pi^a(R)P(d<d_{ub}|R) \nonumber \\
        \text{s.t.}&\qquad p_\pi^a(A)P(d<d_{ub}|A)\geq p_\pi^a(A)(1-\epsilon).
    \end{align}
    Since we are interested in scenarios in which a policy with $p_\pi^a(A)>0$ is feasible, problems (\ref{Eq:opt1}) and (\ref{Eq:opt2}) become similar and maximizing the average reward $r_{\lambda}(\pi)$ with respect to $\pi$ will be the same as computing the Lagrangian dual function associated with problem~(\ref{Eq:opt2}), i.e., 
    \begin{equation}
        \tilde{g}(\lambda)  = \max_{\pi} \tilde{L}(\pi, \lambda) = \max_{\pi} r_{\lambda}(\pi),
        \label{eq:dual2}
    \end{equation}
    where $\tilde{L}(\pi, \lambda)=r_{\lambda}(\pi)$ is the Lagrangian function associated with problem~(\ref{Eq:opt2}). Therefore, $\lambda$ can be seen as a hyper-parameter for our RL problem, where choosing the proper $\lambda$ can result in achieving the goal formulated in (\ref{Eq:opt1}). It should be noted that based on the KKT (Karush–Kuhn–Tucker) conditions, the optimal point $\lambda^*$ must satisfy $\lambda^* \left(P(d<d_{ub}|A)- (1-\epsilon_{ub})\right)=0$.
\end{itemize}

\subsection{Implementation Challenges}\label{challenge}
As mentioned earlier in this section, the immediate reward for a given time step depends on the end-to-end delay of the arriving job at the beginning of the same time step. There are two practical issues regarding this design of the reward function that must be addressed. First, there is no guarantee that the accepted job will finish its end-to-end service by the next time step, and therefore the immediate reward cannot be calculated for the corresponding action taken in that time step until the job has departed. The second practical issue is that the rejected jobs will never go through the network and therefore, the end-to-end delay will not be defined for those jobs. We discuss how these issues can be addressed in this subsection.

\begin{figure}[t!]
\centering
\includegraphics[trim={6cm 7.6cm 6cm 7.6cm},clip, scale=0.5]{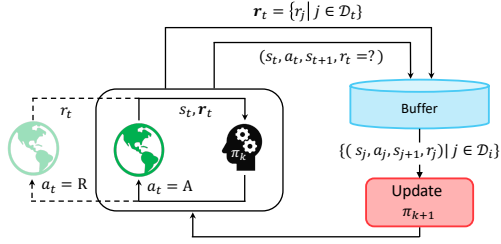}
\caption{Illustration of our reinforcement learning problem. $\mathcal{D}_t$ denotes the set of rewards that correspond to the departed jobs or possibly the  rejected job in time step $t$.}
\vspace{-0.4cm}
\label{fig:env}
\end{figure}

Let us first define $\mathcal{D}_t$ as the set of departed jobs in time step $t$. Therefore, in time step $t$, the end-to-end delay and the immediate reward can be calculated for the jobs in $\mathcal{D}_t$. Hence, instead of updating the Q-function at time step $t$ only based on the experience gained in the current time step, we use the experiences of the departed jobs in time slot $t$, i.e., jobs belonging to $\mathcal{D}_t$. As shown in Fig.~\ref{fig:env}, at any time step $t$, we store the current state $s_t$, the taken action $a_t$ and the next state of the environment $s'_t$ as an incomplete experience tuple, i.e., $(s_t,a_t,s'_t,r_t=?)$, in a buffer. Once the job is departed, the environment returns the corresponding reward $r_t$ and then we can use the complete experience tuple $(s_t,a_t,s'_t,r_t)$ to update the Q-function. Since at time step $t$, we have $|\mathcal{D}_t|$ departures and therefore $|\mathcal{D}_t|$ new complete experience tuples, it is either possible that the Q-function does not get updated in this time step ($|\mathcal{D}_t|=0$), or get updated multiple times ($|\mathcal{D}_t|>1$). It should be noted that the main purpose of using the buffer is to store the incomplete experiences until their corresponding reward becomes available. However, this buffer can also be used as a replay buffer in deep Q-learning settings, where a mini-batch method is used to update the weights by  sampling experiences uniformly from the replay buffer.

Now, let us address the second practical issue regarding the rejected jobs. Since the rejected jobs do not go through the network, the real environment cannot be used to produce the rewards as defined in Eq.~(\ref{Eq:reward}). Instead, we can use a simulated model to generate hypothetical experiences and use them to train the controller. As shown in Fig.\ref{fig:env}, whenever the agent rejects a job, we simulate a parallel environment, shown in grey, with the same state as the original environment upon job arrival, and send the job into the simulated network to measure its end-to-end delay. If the end-to-end delay is smaller than $d_{ub}$, the simulated environment returns a reward of  $r^R_1$, otherwise it returns $r^R_2$. It should be mentioned that only the immediate reward is generated by the simulated environment, while the next state comes from the real environment. Furthermore, since the simulated environment can be fast-forwarded, the end-to-end delay and therefore the corresponding reward can be calculated in the same time step. As a result, we can use the simulated experience to update the policy in the same time step. Algorithm~\ref{algo:ac} shows our R-learning-based algorithm for training the controller.

\begin{algorithm}[]
\SetAlgoLined
 Initialize $\bar{r}$ and $Q(s,a)$
arbitrarily for all $s\in\mathcal{S}$, $a\in\mathcal{A}$

Initialize $s_1$
 
\For{$t=1$ \KwTo \upshape max step}{
\eIf{rand(.)$<\epsilon$}{
 Choose action $a_t$ randomly
 
 $flag_t$=FALSE
}
{$a_t = \arg\max_a Q(s_t,a)$

$flag_t$=TRUE}

Take action $a_t$ and observe $s_{t+1}$ and $\mathbf{r}_t=\{r_i|i\in \mathcal{D}_t\}$

\If{$r_t \notin \mathcal{D}_t$}
{
 Store incomplete experience $(s_t,a_t,s_{t+1},-)$ in the buffer
 
 Store $flag_t$
}
\For{$i$ \textbf{in} $\mathcal{D}_t$}
{Restore $(s_i,a_i,s_{i+1},-)$ from the buffer and append $r_i$

Restore $flag_i$

$\delta_i \leftarrow r_i-\bar{r}+\max_a Q(s_{i+1},a)-Q(s_i,a_i)$

\If{$flag_i=$\text{\upshape TRUE}}{
$\bar{r}\leftarrow \bar{r}+\beta \delta_i$
}}}
\caption{Admission control using R-learning}
\label{algo:ac}
\end{algorithm}
\vspace{-0.4cm}
\section{Evaluation and Results}\label{eval}
\begin{figure}[!t] 
\centering
\subfloat[]{\includegraphics[trim={8.8cm 8.15cm 8.8cm 9.5cm},clip, scale=0.35]{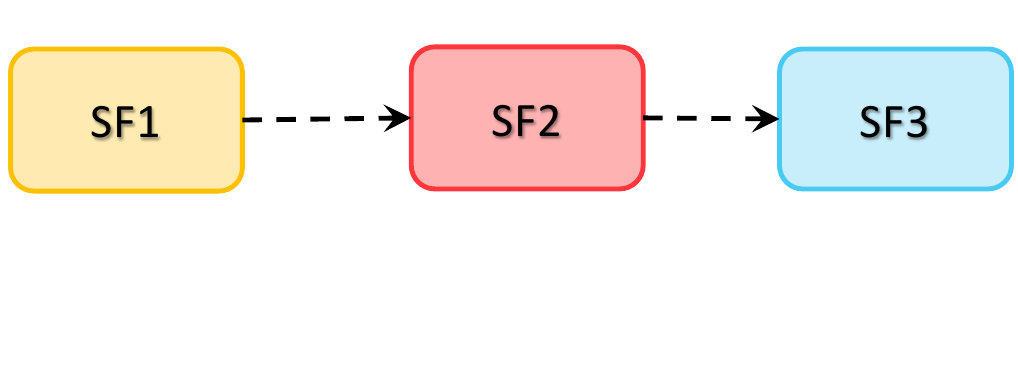}
\label{fig:sfc_tandem}}
\quad
\subfloat[]{\includegraphics[trim={8.8cm 8cm 8.8cm 8cm},clip, scale=0.35]{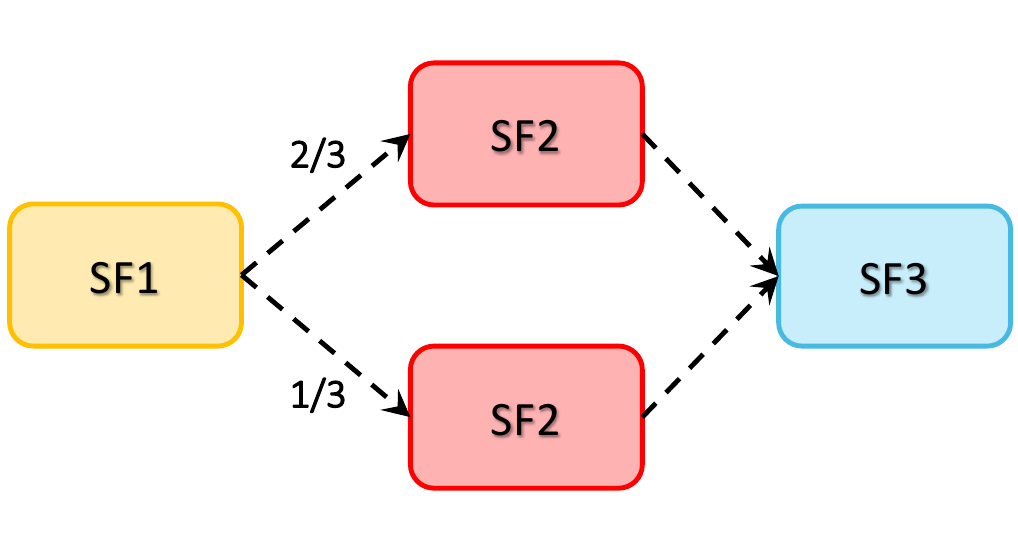}
\label{fig:sfc_acyclic}}
\caption{Service function chain topologies.}
\vspace{-0.5cm}
\end{figure}

As discussed in Section~\ref{intro}, service function chaining is one of the examples of the service networks that can benefit from our RL-based admission controller. In this section, we evaluate the performance of our admission controller in two different scenarios. In the first scenario, we consider a service chain as in Fig.~\ref{fig:sfc_tandem}, where the goal is to provide a probabilistic upper-bound on the end-to-end delay of the accepted jobs. In the second scenario, we consider a service chain with a topology as in Fig.~\ref{fig:sfc_acyclic}, where the application requires that the jobs meet an end-to-end deadline, otherwise they are considered useless.

\begin{figure}[!t] 
\centering
\subfloat[]{\includegraphics[scale=0.35]{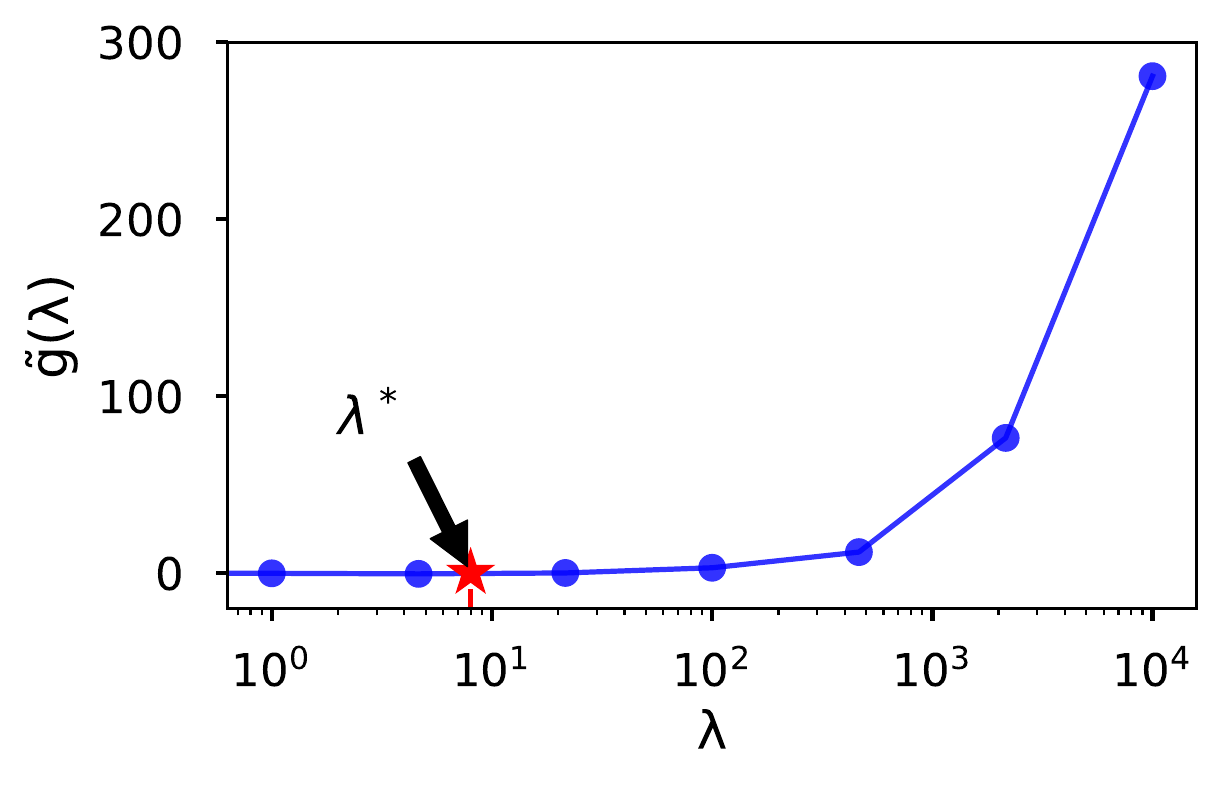}
\label{fig:ave_reward_lambda}}
\subfloat[]{\includegraphics[scale=0.35]{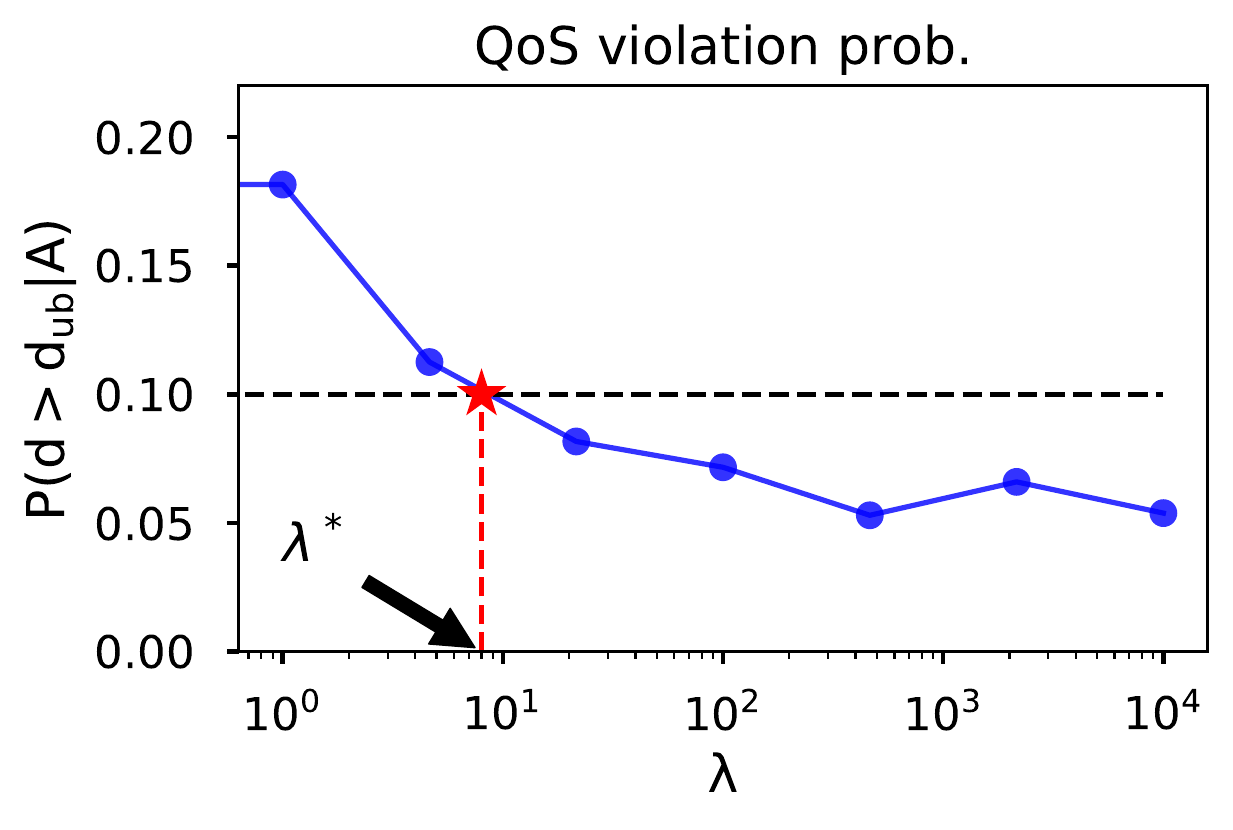}
\label{fig:Prob_violation_lambda}}
\caption{Performance of the admission controller for different values of $\lambda$, a) Maximized average reward as a function of $\lambda$ (Eq. (14)) b) QoS violation probability}
\vspace{-0.5cm}
\end{figure}

Let us start with the first experiment. We model the service function chain with a tandem queueing system as in Fig.~\ref{fig:tandem_topo}, parameters of which are summarized in Table~\ref{ta:sim}.  We assume that the job inter-arrival times and the service times have Gamma distribution (Table~\ref{ta:sim}). In both experiments, time is normalized by the mean service time of the ingress queue, i.e., $1/(c_1 \mu_1)$. Our goal is to design an admission controller that provides an upper-bound of $d_{ub}=15$ on the end-to-end delay of the jobs, with violation probability $\epsilon_{ub}=0.1$, and minimum unnecessary job rejections. In order to achieve this goal, we use our proposed RL-based admission controller. As discussed in section~\ref{AC_RL}, the reward function parameters in Eq.~(\ref{Eq:reward}) are set to $r_1^A = \epsilon_{ub} \lambda$, $r_2^A = -(1-\epsilon_{ub})\lambda$, $r_1^R = -1$ and $r_2^R = 0$, where $\lambda=\lambda^*$ is the optimal solution to Eq.~(\ref{eq:dual2}).
As mentioned earlier, the optimal value of hyper-parameter $\lambda$, i.e., $\lambda^*$, must satisfy $\lambda^* \left(P(d<d_{ub}|A)- (1-\epsilon_{ub})\right)=0$, where $\lambda^*\geq0$. Therefore, in the optimal point either $\lambda^*=0$ or $P(d>d_{ub}|A)=\epsilon_{ub}$. Fig.~\ref{fig:ave_reward_lambda} shows the optimized average reward, i.e. $\tilde{g}(\lambda)=\max_{\pi}r_\lambda(\pi)$, as a function of $\lambda$. Based on Eq.~(\ref{eq:dual2}), $\lambda^*$ can be obtained by finding the $\lambda$ for which the optimized average reward function, i.e. the Lagrangian dual function $\tilde{g}(\lambda)$, is minimized. As shown in Fig.~\ref{fig:ave_reward_lambda}, the minimum is achieved for $\lambda^*=8$. This can also be verified using Fig.~\ref{fig:Prob_violation_lambda}, in which $\lambda^*$ satisfies the QoS constraint.
\begin{figure*}[!t] 
\centering
\subfloat[]{\includegraphics[width=.29\linewidth, height=3cm]{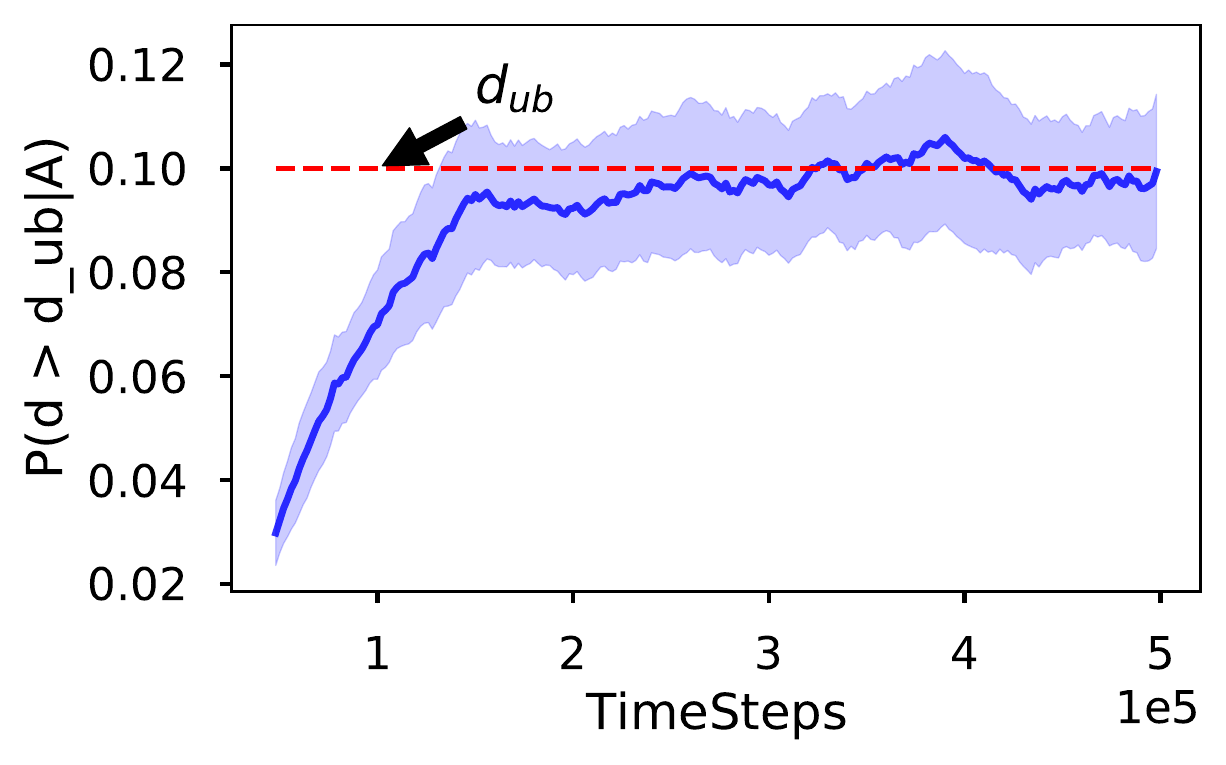}
\label{fig:qos}}
\subfloat[]{\includegraphics[width=.29\linewidth, height=3cm]{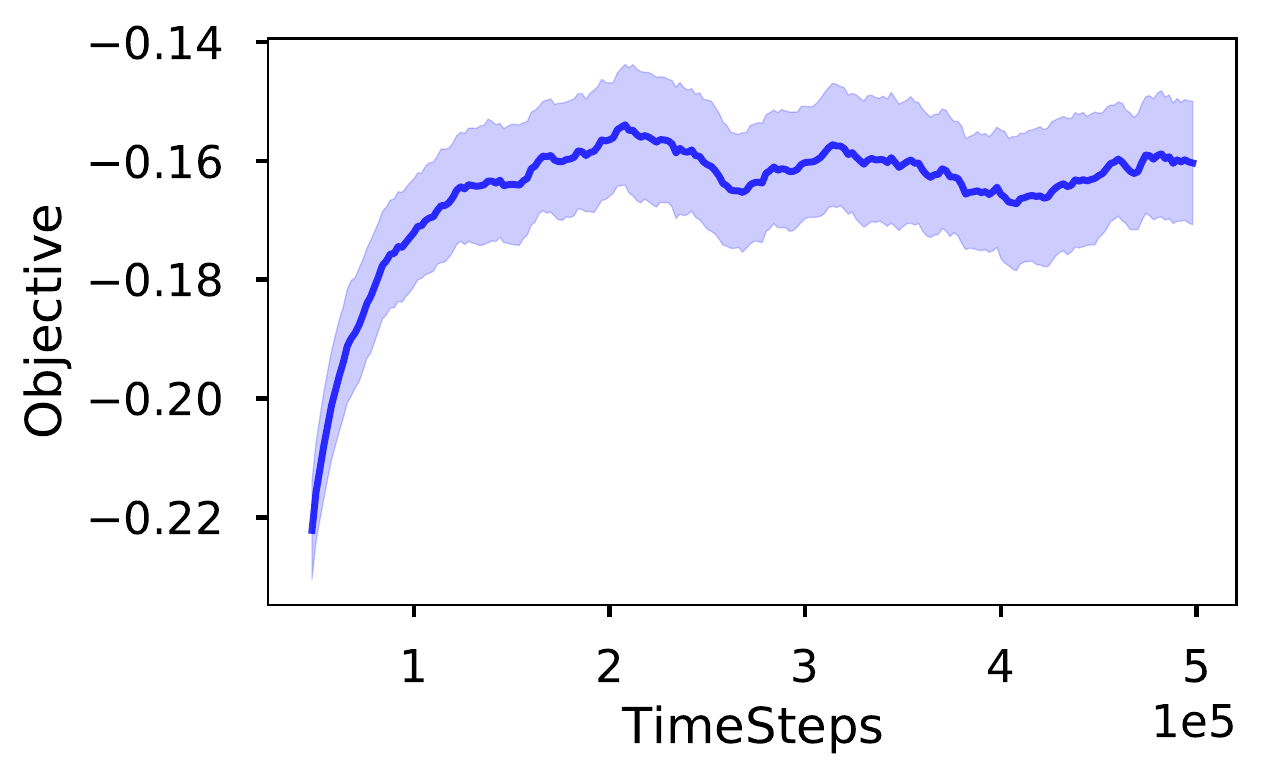}
\label{fig:objective}}
\subfloat[]{\includegraphics[width=.29\linewidth, height=3cm]{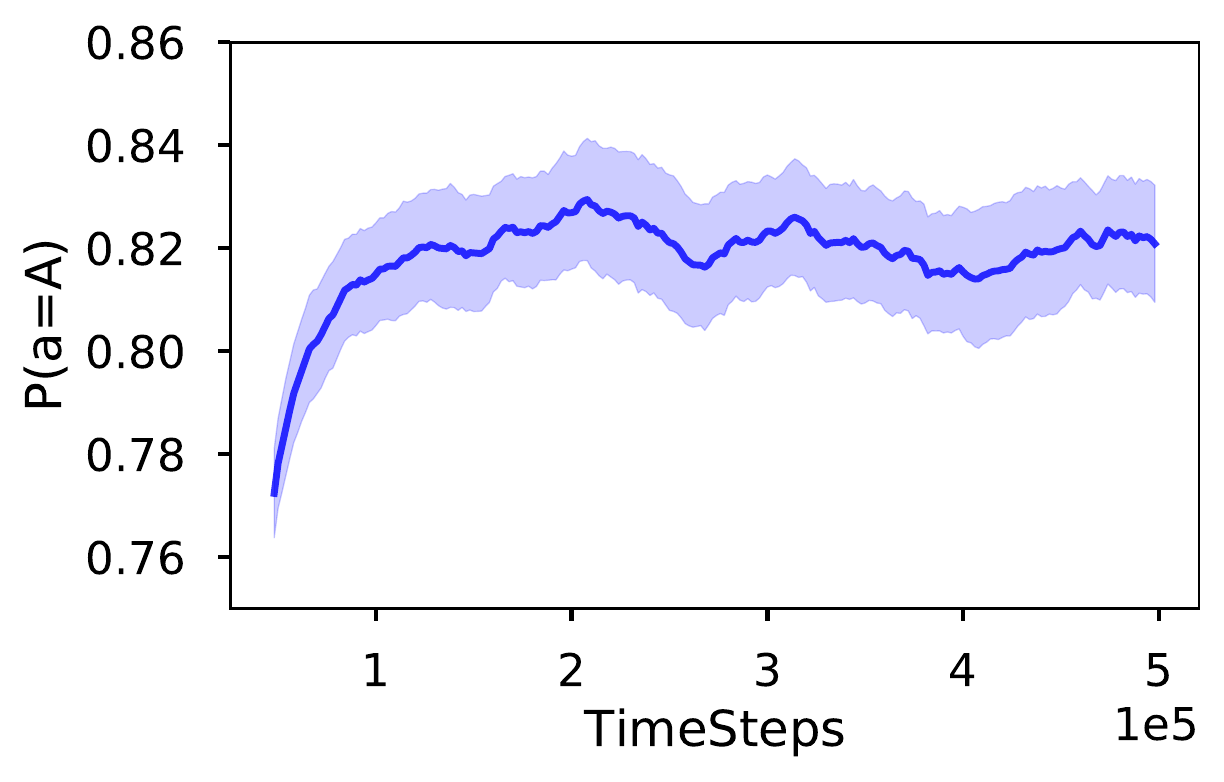}
\label{fig:acceptance}}\hfill
\subfloat[]{\includegraphics[width=.29\linewidth, height=3cm]{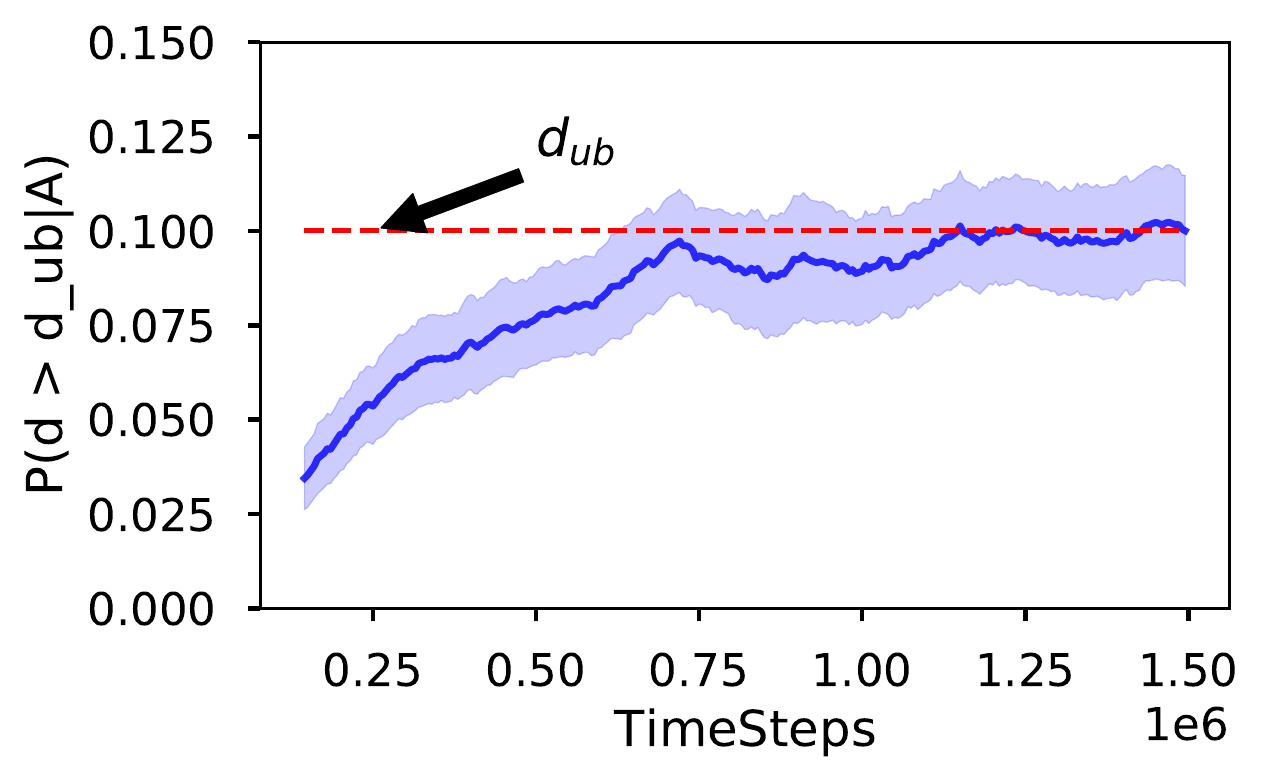}
\label{fig:qos_acyc}}
\subfloat[]{\includegraphics[width=.29\linewidth, height=3cm]{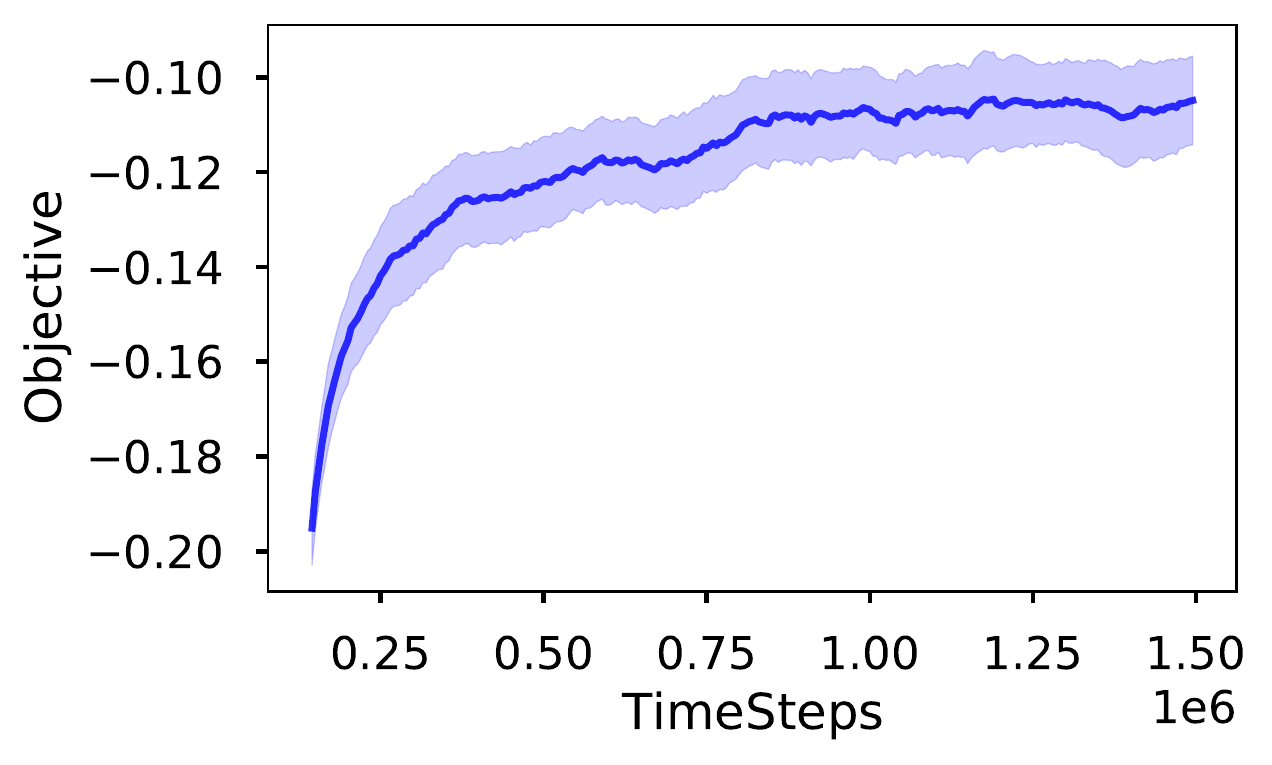}
\label{fig:objective_acyc}}
\subfloat[]{\includegraphics[width=.29\linewidth, height=3cm]{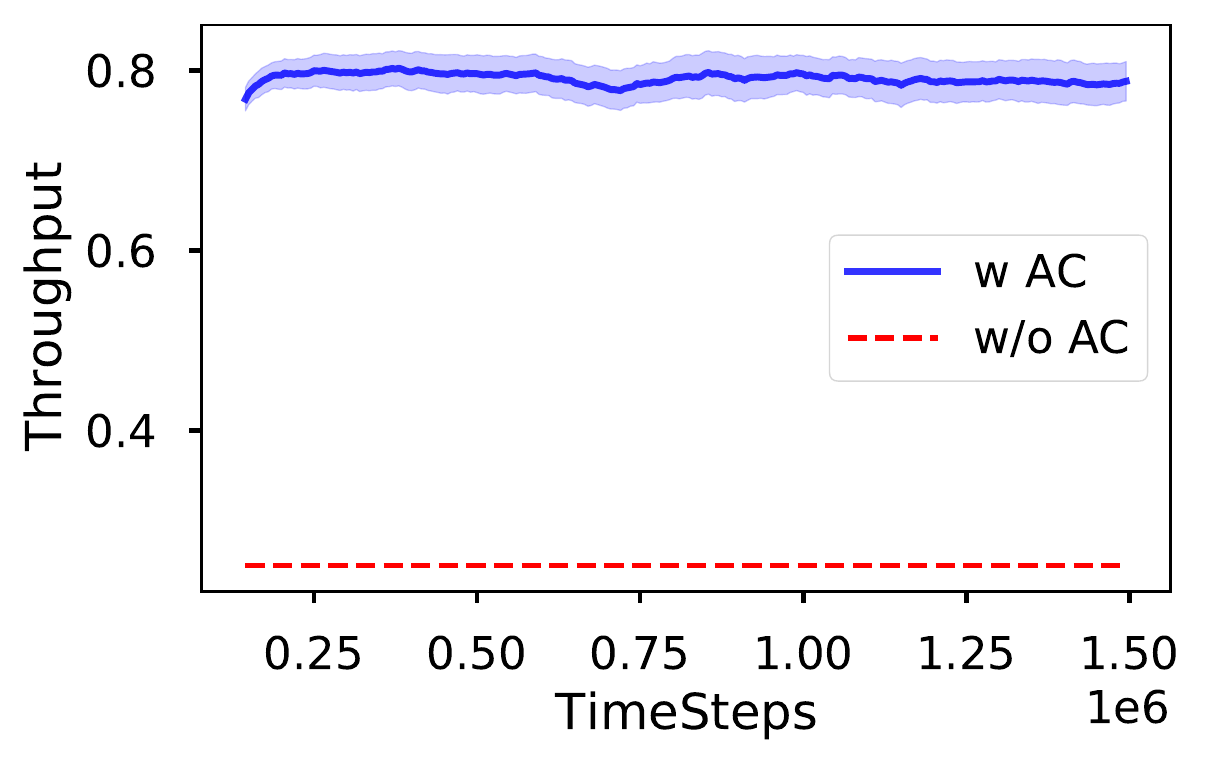}
\label{fig:throughput_acyc}}
\caption{Performance of the RL-based admission controller in: scenario~I: a) QoS constraint b) objective function c) acceptance rate; scenario~II: a) QoS constraint b) objective function c) throughput}
\label{fig:learning_curves}
\vspace{-0.5cm}
\end{figure*}
\begin{table}[!t]
\caption{Simulation Parameters}
\centering
\begin{tabular}[!t]{lll}
\hline
\rowcolor{Gray}
\bfseries Topology& \bfseries \parbox{5mm}{Num.~of~servers\\$[c_1,c_2, \cdots, c_N]$} & \bfseries \parbox{5mm}{Service~rates\\$[\mu_1,\mu_2, \cdots, \mu_N]$} \\ 
\hline
\hline
Tandem &$[3,5,2]$ & $[0.33,0.2,0.5]$\\ 
Acyclic &$[5,3,3,2]$ & $[0.2,0.22,0.11,0.5]$\\
\hline
\rowcolor{Gray}
\bfseries Distribution& \bfseries Parameters & \\ 
\hline
\hline
Gamma (Arrival)&$\lambda = 0.95$, SCV = $0.7$\\ 
\hline
Gamma (Service time)& SCV = $0.8$\\ 
\end{tabular}\label{ta:sim}
\vspace{-0.5cm}
\end{table}

Now, using $\lambda^*$, the admission controller, i.e., the agent, has been trained with 4 different initial seeds using Algorithm~\ref{algo:ac}. Fig.\ref{fig:learning_curves}a-c show the performance of our trained controller in scenario I. The dark blue curves and the pale blue regions show the average and the standard error bands, respectively. Fig.~\ref{fig:qos} shows that the QoS constraint is satisfied, i.e., $P(d>d_{ub}|A)$ converges to $\epsilon_{up}=0.1$ as we train the agent. This should be expected in the optimal point, since $\lambda^*>0$. On the other hand, Fig.~\ref{fig:objective} shows the maximization of the objective function, which is equal to minimizing the average number of unnecessary rejections per time step. As can be seen in Fig.~\ref{fig:acceptance}, the acceptance rate of the trained admission controller, which is equal to the throughput of the system in this scenario, converges to around $82\%$.

In the second experiment, we consider a service function chain as in Fig.~\ref{fig:sfc_acyclic}, where service function SF2 has two instances with separate physical resources. Therefore, we can model the system by an acyclic network as in Fig.~\ref{fig:acyclic_topo}, with two parallel branches. The parameters of the model are summarized in Table~\ref{ta:sim}. We assume that the application imposes a deadline of $d_{ub}=20$ on the end-to-end delay of the jobs. Using our proposed method, we can train an admission controller that guarantees that the accepted jobs meet the deadline with probability $1-\epsilon_{ub}=0.9$. We can use a similar approach as in the previous experiment to find $\lambda^*$, which will be equal to $\lambda^*=5$ in this setting. As shown in Fig.~\ref{fig:qos_acyc}, the probability that an accepted job fails the end-to-end delay deadline converges to $\epsilon_{ub}=0.1$. Moreover, Fig.~\ref{fig:throughput_acyc} shows that the admission controller has tremendously improved the throughput of the service function chain, compared to the case with no admission controller.

\vspace{-0.1cm}
\section{Conclusions}\label{con}
    In this paper, we proposed an RL-based admission controller for providing end-to-end delay guarantees in service networks. We adopted an average-reward reinforcement learning approach to handle the infinite horizon problem. The reward function is designed such that maximizing the average reward results in minimizing the probability of unnecessary job rejections, conditioned on satisfying the QoS constraint. The admission controller only observes the queue length values and does not require any information about the network topology or other system parameters.

\end{document}